\documentclass[aps,twocolumn,a4paper,showpacs]{revtex4}
\usepackage{graphicx}
\usepackage{amsmath}
\usepackage{amssymb}
\usepackage{enumerate}
\usepackage{subfigure}
\usepackage{tabularx}

\newcommand{\be}{\begin{equation}}
\newcommand{\ee}{\end{equation}}
\newcommand{\ben}{\begin{eqnarray}}
\newcommand{\een}{\end{eqnarray}}
\newcommand{\bes}{\begin{subequations}}
\newcommand{\ees}{\end{subequations}}

\newcommand{\bb}{\bibitem}
\newcommand{\sech}{{\rm sech}}

\begin{document}
\title{From Kinks to Compactons}
\author{D. Bazeia$^{1,2}$, L. Losano$^{1,2}$, M.A. Marques$^1$, and R. Menezes$^{2,3}$}
\affiliation{Departamento de F\'\i sica, Universidade Federal da Para\'\i ba, 58051-970 Jo\~ao Pessoa, PB, Brazil}
\affiliation{Departamento de F\'\i sica, Universidade Federal de Campina Grande, 58109-970, Campina Grande, PB, Brazil}
\affiliation{Departamento de Ci\^encias Exatas, Universidade Federal
da Para\'{\i}ba, 58297-000 Rio Tinto, PB, Brazil}

\begin{abstract}
This work deals with the presence of localized structures in relativistic systems described by a single real scalar field in two-dimensional spacetime. We concentrate on kinks and compactons in models with standard kinematics, and we develop a procedure that help us to smoothly go from kinks to compactons in the suggested scenario. We also show how the procedure works in the braneworld scenario, for flat brane in the five-dimensional spacetime with a single extra dimension of infinite extent. The brane unveils a hybrid profile when the kink becomes a compacton, behaving as a thick or thin brane, depending on the extra dimension being inside or outside a compact space.
\end{abstract}
\date{\today}
\pacs{11.27.+d, 11.10.Lm}
\maketitle


{\it Introduction.} Topological structures appear in high energy physics in a diversity of contexts \cite{b1,b2}. The standard topological structures are static solutions of the equations of motion, known as kinks, vortices and monopoles. Kinks are the simplest structures, and appear in the presence of scalar fields in $(1,1)$ spacetime dimensions. Because of its intrinsic simplicity, kinks can be used to describe the behavior of physical systems in several branches of physics; see, e.g., Ref.~\cite{e1,e2,e3,e4,e5,e6,e7,e8}. For instance, kinklike structures may be used to model solutions for the one dimensional Bogoliubov-de Gennes equations and/or Eilenberger equation of semiclassical superconductivity \cite{e2}, the presence of Bose-Einstein condensates with two- or three-body interactions \cite{e3}, moduli space of non-relativistic waves in the long wavelength limit of ferromagnetic spin chains \cite{e5}, the process of production of kink-antikink pairs in the collision of particle-like states \cite{e8}, etc. Kinks may also be used to model the fifth dimension, in braneworld models with a single extra dimension of infinite extent; see, e.g., Refs.~\cite{RS,GW,MC,brane}. 

In this work we focus attention on kinks, so we deal with scalar fields in two-dimensional spacetime. Our main interest is to develop a smooth  connection between kinks and compactons, which are kinks that live in a compact space. While a kink is a linear structure with energy density that vanishes asymptotically, a compact kink is a linear compact structure, with energy density that vanishes outside a closed interval of the real line.

Compactons first appeared in \cite{prl}, in models comprising nonlinearity and nonlinear dispersion. However, since there is no way to include nonlinear dispersion in relativistic systems governed by standard kinematics, compactons seem to be very hard to appear in standard field theory. More recently, however, some authors have investigated the presence of compactons in models with standard kinematics \cite{s1,s2,s3,s4}. In particular, the investigations \cite{s3,s4} have inspired us to develop a procedure to smoothly navigate from kinks to compactons, shedding light on the game that compact kinks may play in high energy physics. Moreover, we also investigate how the procedure works in the braneworld scenario, with an $AdS_5$ warped geometry with a single extra dimension of infinite extent.

{\it Generalities.} In order to investigate the problem, let us write the Lagrange density for a scalar field with standard kinematics. It has the form
\be\label{model}
{\cal L}=\frac12\partial_\mu\phi\partial^\mu\phi -V(\phi).
\ee
The metric is $(+,-)$, $\mu=0,1$, and $\phi$ is a real scalar field. $V(\phi)$ is the potential and identifies the way the scalar field
self-interacts. The equation of motion is given by
\be\label{eqMotion1}
\frac{d^2 \phi}{dt^2}-\frac{d^2 \phi}{dx^2}  + \frac{dV}{d\phi}=0.
\ee
We consider a non negative potential $V(\phi)$ that supports the set of minima $\{v_i, i=1,2,...\},$ with $dV/d\phi|_{v_i}=0$ and $V(v_i)=0$.  A topological sector is characterized by two neighbor minima, and the topological solution is a kink, $\phi_k(x)$, that connects the minima $v_k$ and $v_{k+1}$. It obeys the equation
\be
\frac{d^2\phi_k}{dx^2}=\frac{dV}{d\phi}\bigg|_{\phi_k},
\ee
with $\phi_k(x\to -\infty)\to v_k$ and $\phi_k(x\to\infty)\to v_{k+1}$, with vanishing derivatives.
It has amplitude $a_k=|v_{k}-v_{k+1}|$ and energy density $\rho_k(x)$, given by 
\be
\rho_k(x)\!=\!\frac12 \left(\!\frac{d\phi_k}{dx}\!\right)^2\!\!\!+V(\phi_k(x))\!=\!\left(\!\frac{d\phi_k}{dx}\!\right)^2\!\!\!=\!2V(\phi_k(x)).
\ee

We study linear stability, to see how the solution behaves under the presence of small fluctuations. We add small fluctuations around the static solution $\phi_k(x)$, writing $\phi(x)=\phi_k(x)+\eta(x)\cos(\omega t)$. We use this into the equation of motion and expand it up to first-order in $\eta$ to get the Schroedinger-like equation 
\be \label{stab}
\left(-\frac{d^2}{dx^2}+U_k(x) \right)\eta=\omega^2 \eta,
\ee
with 
\be\label{qm}
U_k(x)=\left.\frac{d^2V}{d\phi^2}\right|_{\phi=\phi_k(x)}.
\ee
This is the potential that appears in the study of statility, so we call it the stability potential. 
We see from the above Eq.~\eqref{qm} that $U_k(x)$ goes asymptotically to $m^2_k$ and $m^2_{k+1}$, which are the (squared) masses of the elementary excitations at the minima $v_k$ and $v_{k+1}$, respectively.

An important model which falls within the class of models that we are interested in is the $\phi^4$ model, with spontaneous symmetry breaking. The potential is usually given by $\lambda^2(v^2-\phi^2)^2/2$. However, in 
$(1,1)$ dimensions, the field $\phi$ and the parameter $v$ are dimensionless; so, $\partial_\mu\phi$ and $\lambda$ have dimension of energy. We redefine $\phi\to v\phi$ and $\partial_\mu\phi\to\lambda v^2\partial_\mu\phi$, such that
${\cal L}=\lambda^2 v^4{\cal L'}$. We omit the prime to get the Lagrange density ${\cal L}^\prime$ as in \eqref{model}, with
\be\label{p4}
V(\phi)=\frac12(1-\phi^2)^2.
\ee
Here we are considering the field $\phi$ and the coordinates $x$ and $t$ dimensionless. We focus on this model, which has the minima
$v_{\pm}=\pm 1$ and a maximum at the origin, such that $V(0)=1/2$. We also have $m^2_-=m^2_+=m^2=4$, and the kink solution is given explicitly by
\be\label{k}
\phi(x)= \tanh( x),
\ee
where we are taking the center of the kink at the origin, $x=0$, for simplicity. The energy density has the form
\be\label{ed}
\rho=\sech^4(x),
\ee
and the energy is $E=4/3$. In this case, the stability potential of Eq.~\eqref{qm} is given by
\be
U(x)=4-6 \,{\rm sech}^2(x).
\ee
It is the modifield Poeschl-Teller potential \cite{pt}; it is reflectionless and has two bound states: the zero mode, with $\omega_0=0$, and one excited state, with $\omega^2_1=3$.

Let us now investigate how a compact kink appears in a relativistic model. We introduce the Lagrange density
\be
{\cal L}=-\frac14(\partial_\mu\phi\partial^\mu\phi)^2-\frac32 V(\phi),
\ee
where $V(\phi)$ is given by \eqref{p4}. This model has generalized kinematics, and the equation of motion for static field is now given by
\be
\left(\frac{d\phi}{dx}\right)^2\frac{d^2\phi}{dx^2}=-\phi(1-\phi^2).
\ee
It engenders nonlinearity, which comes from the potential, and nonlinear dispersion, which enters the game from the generalized kinematics introduced in the model. This is the desired picture, as it is suggested in the original work on compactons \cite{prl}. See Refs.~\cite{babi,blmo} for more details on this. 

The compact kink has the form
\begin{equation}\label{ck}
\phi(x)=\left\{
\begin{array}{ll}
1 &  \mbox{ for } x> \pi/2;\\ 
\sin(x)& \mbox{ for } |x|\leq\pi/2; \\ 
-1  & \mbox{ for } x < -\pi/2.
\end{array} \right.
\end{equation}
The energy density is given by 
\be 
\rho_c(x)=
\begin{cases}
\cos^4(x),\,\,\,& |x|\leq\pi/2;\\
0,\,\,\,&|x|>\pi/2.\end{cases}
\ee
The energy is $E_c=3\pi/8$.

The compact kink \eqref{ck} is stable. The study of stability is similar to the case of kinks, but now the stability potential of \eqref{qm} becomes the Poeschl-Teller potential \cite{pt}, which only supports bound states; see, e.g., Ref.~\cite{blmo} for further details on this.

{\it The connection.} It is hard to see how to go from kinks to compactons within the above two very distinct scenarios. However, in \cite{s3,s4} one has found compact solutions in models with standard dynamics, and this has encouraged us to search for a mechanism to smoothly transform kinks into compactons.
 To go further into this, we remember that the energy density of the kink \eqref{ed} goes to zero asymptotically in the form $\exp{(-4x)}$, which depends on the masss of the elementary excitation. Thus, to make the kink compact, one has to increase the mass to larger and larger values. In fact, we note from the potential that controls the kink fluctuations, given by \eqref{qm}, that if the mass increases to larger and larger values, the potential changes, tending to behave as an infinite well, then localizing the fluctuations into a compact space.
 
For this reason, if we want to go from kinks to compactons, we have to modify the potential in a way such that the associated squared mass may increase to larger and larger values. We have found two distinct possibilities, and we investigate them below.

We first introduce the model
\be
{\cal L}_\alpha =\frac12\partial_\mu\phi\partial ^\mu\phi-V_\alpha(\phi),
\ee
with the potential
\be\label{pot1}
V_{\alpha}(\phi)=\frac{1}{2\alpha}\left(\sqrt{1+4\alpha(1+ \alpha/2)V(\phi)}-1\right).
\ee
Here $\alpha$ is a non-negative real parameter and $V(\phi)$ is the $\phi^4$ potential given by Eq.~\eqref{p4}; it is a non-negative potential which does not depend on $\alpha$. We have, for the equation of motion,
\be\label{eom1}
\frac{d^2\phi}{dx^2}=\frac{1+ \alpha/2 }{\sqrt{1+4\alpha(1+\alpha/2) \; V(\phi)}}\frac{dV}{d\phi}.
\ee
Since the right hand side is the derivative of $V_\alpha(\phi)$, we see that the extrema of $V(\phi)$ are also extrema of $V_\alpha(\phi)$. We also have
\ben
\frac{d^2 V_\alpha}{d\phi^2} &=& \frac{(1+\alpha/2)}{[1+4\alpha(1+\alpha/2)V]^{1/2}}\,\frac{d^2 V}{d\phi^2}\nonumber \\
&& -\frac{2\alpha(1+\alpha/2)^2}{[1+4\alpha(1+\alpha/2)V]^{3/2}}\left(\frac{dV}{d\phi}\right)^2.
\een
Thus, at the minima $v_{\pm}$ we get
\be \label{tildem}
\frac{d^2V_\alpha}{d\phi^2}\bigg{|}_{v_\pm}=(1+ \alpha/2) \frac{d^2V}{d\phi^2}\bigg{|}_{v_\pm},
\ee
which connects the masses of the elementary excitations of the two models. We can write $m^2_\alpha=4+2\alpha$,
which shows that $m_\alpha$ increases smoothly to very large values as $\alpha$ increases to very large values. This is the desired property of the potential, and we now proceed to show how the kink behaves, changing to the desired compact structure.

For $\alpha$ very small we get  
\be
V^s_{\alpha}(\phi)= V(\phi)+\frac12\,\alpha V(\phi) (1-2V(\phi)) +{\cal O}[\alpha^2].
\ee
For $1/\alpha$ very small we get
\be
V^l_{\alpha}(\phi)= \sqrt{V(\phi)/2} + \frac{1}{2\alpha}\left( \sqrt{2V(\phi)}-1\right) + {\cal O}\left[\frac{1}{\alpha^2}\right].
\ee

\begin{figure}[t]
\includegraphics[width=4.2cm]{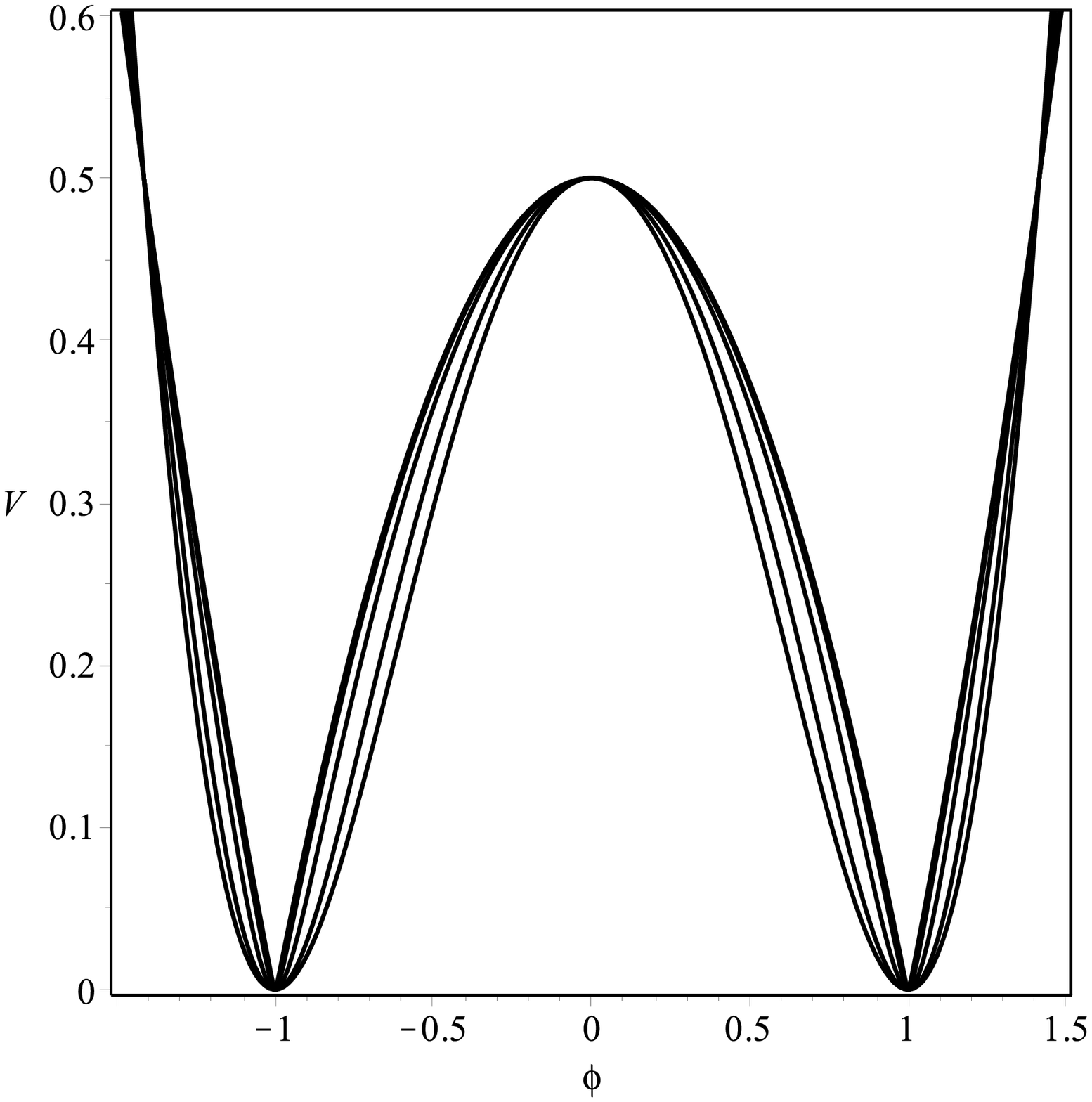}
\includegraphics[width=4.2cm]{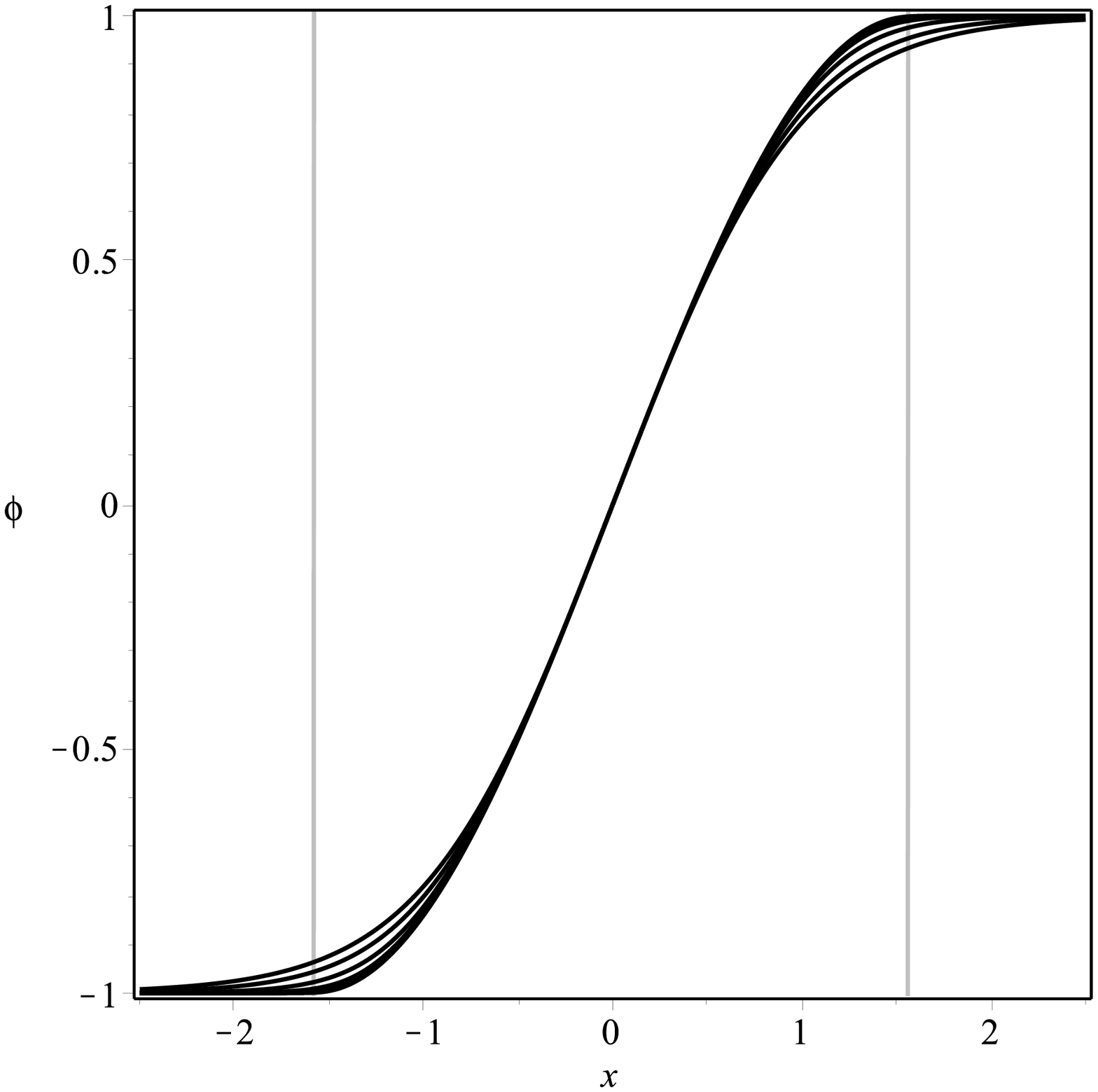}
\includegraphics[width=4.2cm]{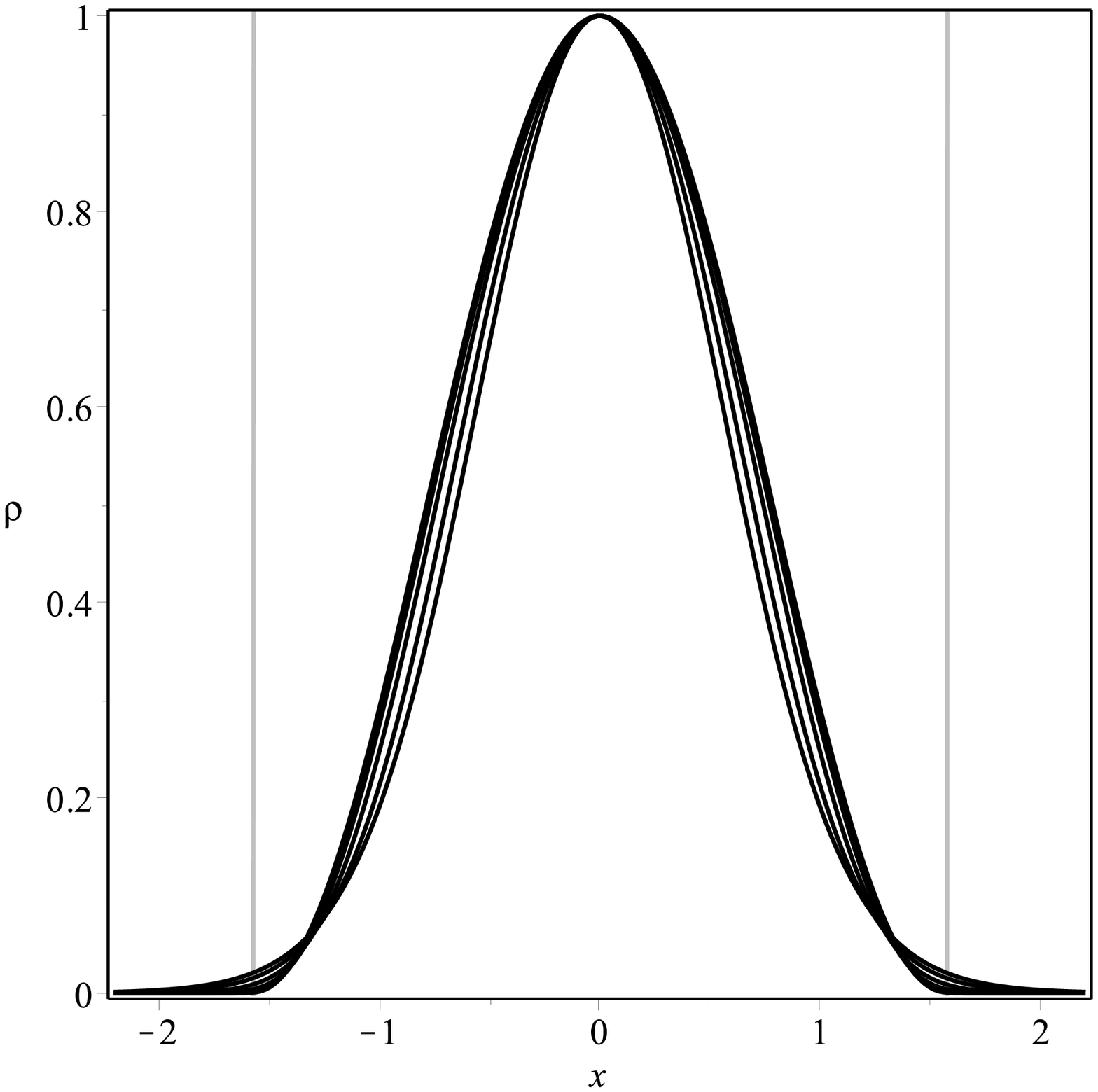}
\includegraphics[width=4.2cm]{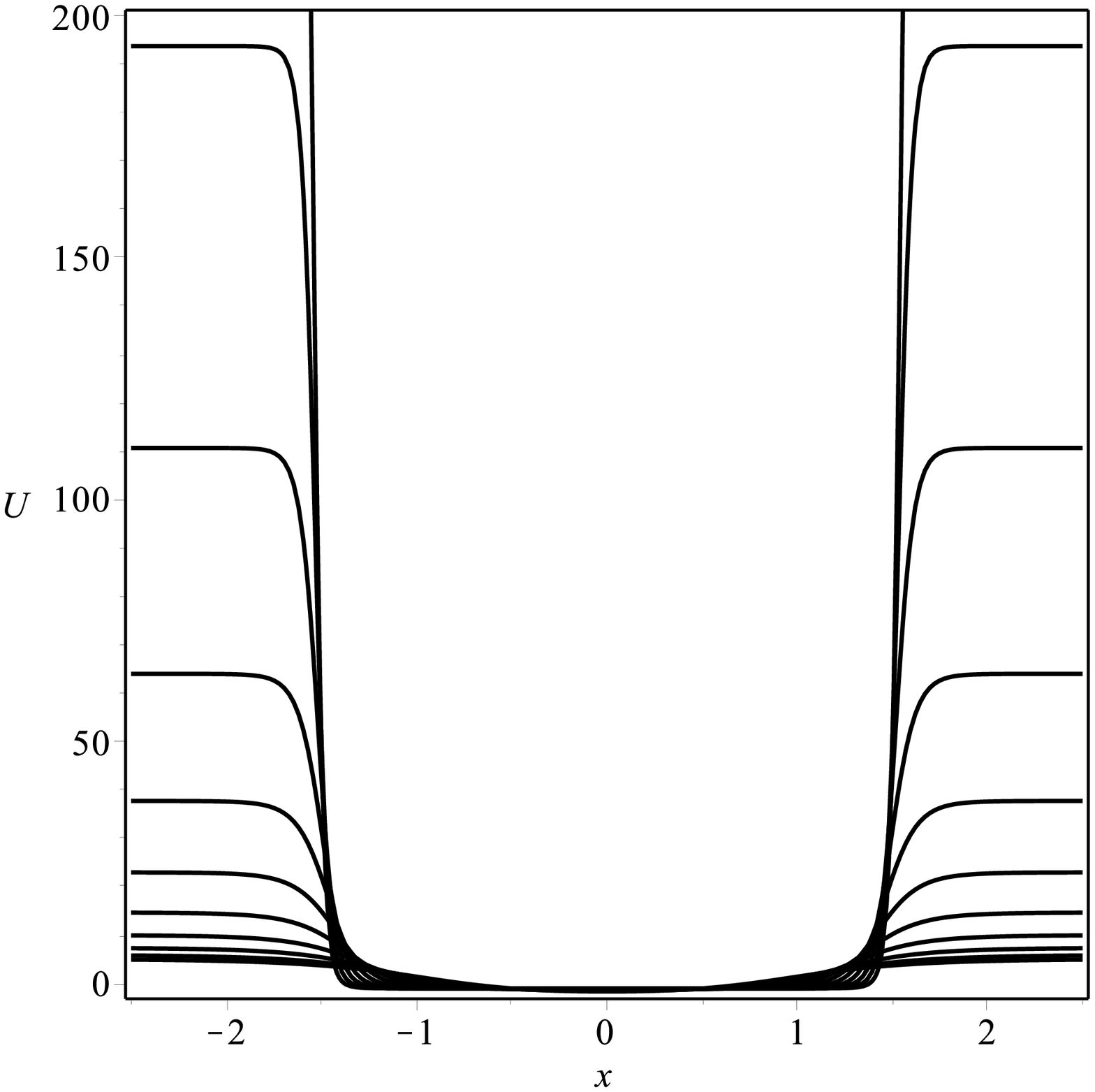}
\caption{The potential \eqref{pot1} (top left), kink solution (top right), energy density (bottom left) and stability potential (bottom right), depicted for $\alpha=0$, and increasing to larger and larger values.}
\label{fig1}
\end{figure}

In the limit $\alpha \to 0$ we get to $V(\phi)$, and we take this as the limit for the standard kink. According to our suggestion, we then take the limit $1/\alpha\to 0$ as the limit for the compact structure. Here we have
\be
V_c(\phi)=\sqrt{V(\phi)/2 },
\ee
as the compacton limit. In this limit, the potential becomes
\be\label{modulus}
 V_c(\phi)= \frac12 |1-\phi^2|,
\ee
In this model with standard kinematics, the compact kink has the very same form given by \eqref{ck}. However, the energy density is different, and here it has the form
\be
\rho_c(x)=
\begin{cases}
\cos^2(x),\,\,\,&|x|\leq \pi/2;\\
0, \,\,\, & |x|>\pi/2,
\end{cases}
\ee 
such that the energy is $E^1_c=\pi/2$.

\begin{figure}[t]
\includegraphics[width=7.2cm]{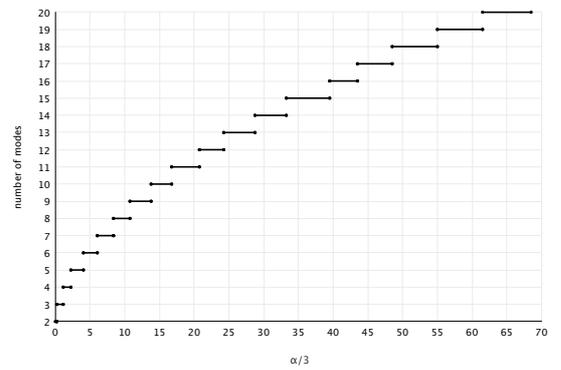}
\caption{Number of bound states of the potential \eqref{} of the first model, depicted as a function of $\alpha$.}
\label{fig2}
\end{figure}
\begin{figure}[t]
\includegraphics[width=4.2cm]{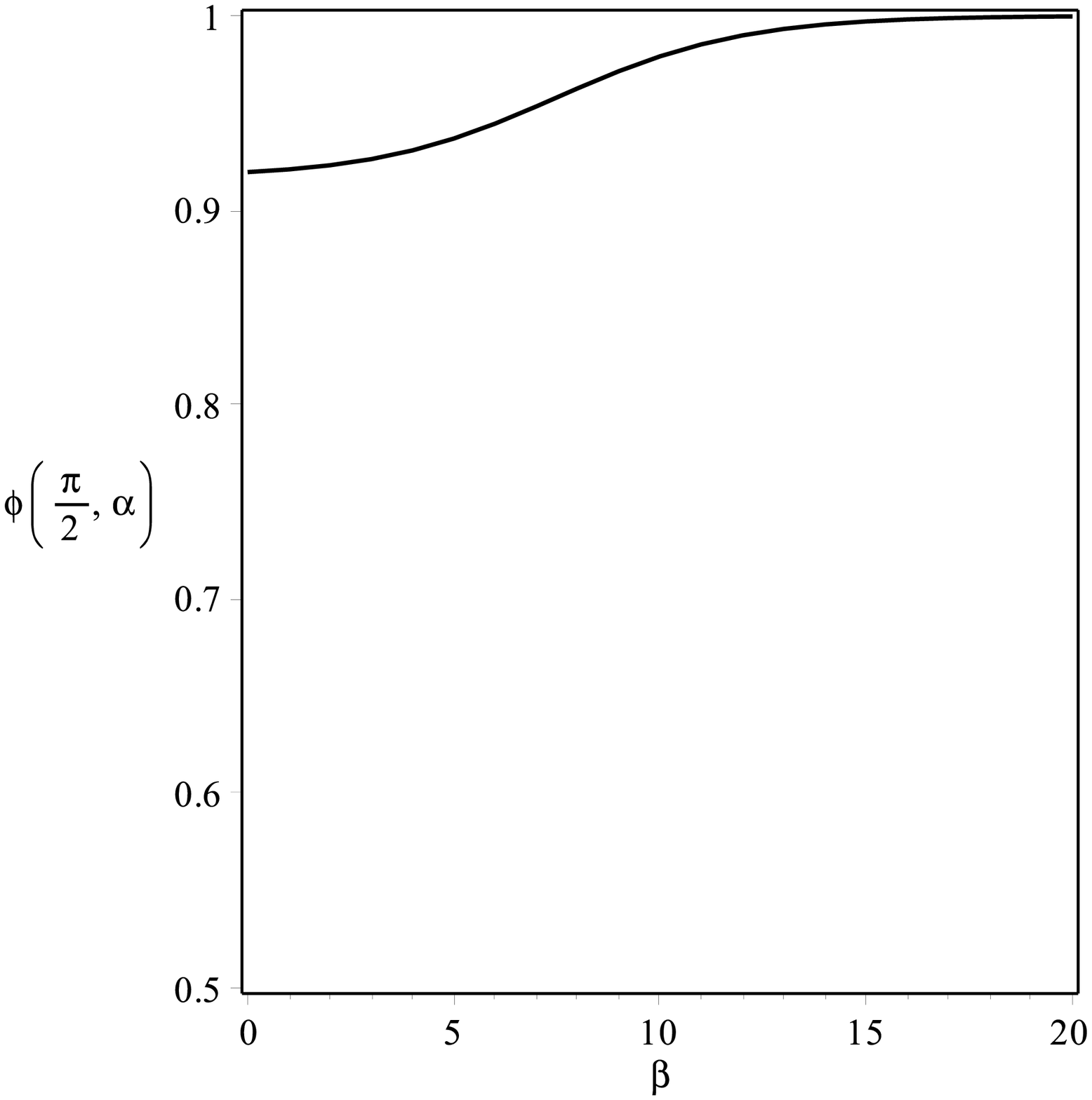}
\includegraphics[width=4.2cm]{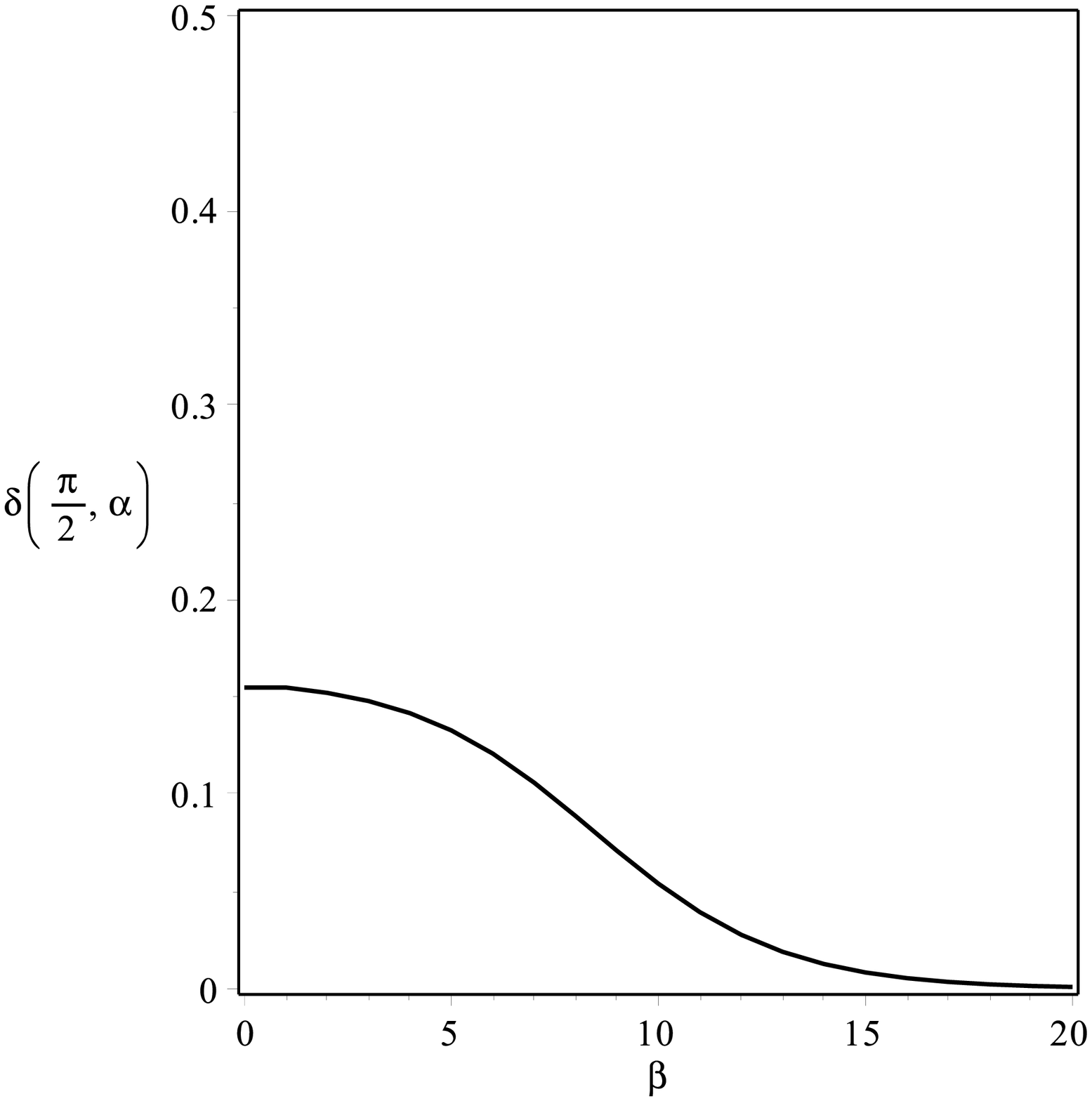}
\caption{Behavior of $\phi(\pi/2,\alpha)$, the kink solution $\phi(x)$ calculated at $x=\pi/2$ (left) and the relative energy $\delta(\pi/2,\alpha)$ of the kink solution (right), for several values of $\beta=5.66+2.13 \ln\alpha$.}
\label{fig3}
\end{figure}

In the compact limit, stablility potential has the form
\be\label{qmc}
U_c(x) =
\begin{cases}
-1, \,\,\, & |x|<\pi/2;\\
\infty,\,\,\,&|x|\geq \pi/2.
\end{cases}
\ee
It is an infinite well, and supports an infinite set of bound states with energies $\omega_k^2=k(k+2)$, where $k=0,1,2,\ldots$.

In order to understand how we smoothly go from kinks to compactons is this case, let us consider the equation of motion \eqref{eom1} using the $\phi^4$ potential. It has the form
\be\label{eom}
\frac{d^2\phi}{dx^2}=-\frac{(2+ \alpha )\phi(1-\phi^2)}{\sqrt{1+\alpha(2+\alpha)(1-\phi^2)^2}}\,.
\ee  
We study this equation numerically, and in Fig.~{\ref{fig1}} we depict the potential, the kink solution of the above Eq.~\eqref{eom},
the energy density and the stability potential for several values of $\alpha$, including
$\alpha=0$, and other larger and larger values.

Let us now focus on the stability potential, to see how is changes with increasing $\alpha$. It is depicted in Fig.~{\eqref{fig1}}, bottom right. It is a well with increasing walls that converge to the form \eqref{qmc} for very large values of $\alpha$. We also depicted in Fig.~\eqref{fig2} the number of bound states as a function of $\alpha$. It shows that the well becomes deeper and deeper, adding more and more bound states, as $\alpha$ increases to larger and larger values.

To make the investigation stronger, in Fig.~3 we study two other quantities numerically, one being $\phi(\pi/2,\alpha)$, that is, the value of the kink solution $\phi(x)$ at the point $x=\pi/2$, as a function of $\alpha$. The other issue concerns the quantity of energy inside the interval $[-\pi/2,\pi/2]$. We studied the quantity $\delta(\pi/2,\alpha)=1-E(\pi/2,\alpha)/E^1_c$, where $E^1_c$ is the energy of the compact kink of the first model, and $E(\pi/2,\alpha)$ is the quantity of energy of the kink inside the interval $|x|\leq\pi/2$. The results show that the kink solution goes to the unit value at $x=\pi/2$, for increasing values of $\alpha$, and that the quantity of energy of the kink inside the compact interval $|x|\leq\pi/2$ goes exactly to the energy of the compact solution. These results confirm that the kink becomes a compact structure as $\alpha$ increases to a very large value. 

\begin{figure}[t]
\includegraphics[width=4.2cm]{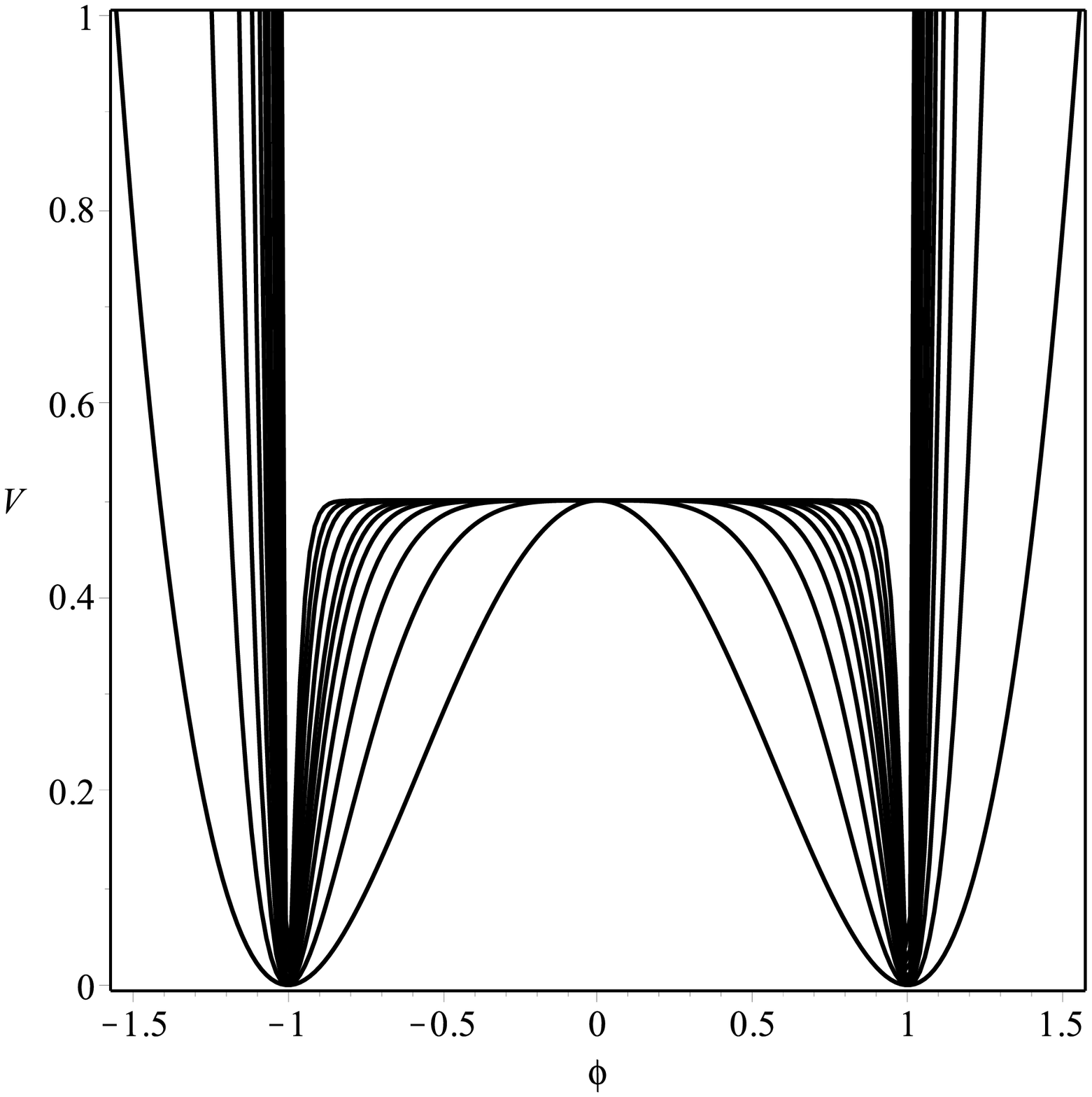}
\includegraphics[width=4.2cm]{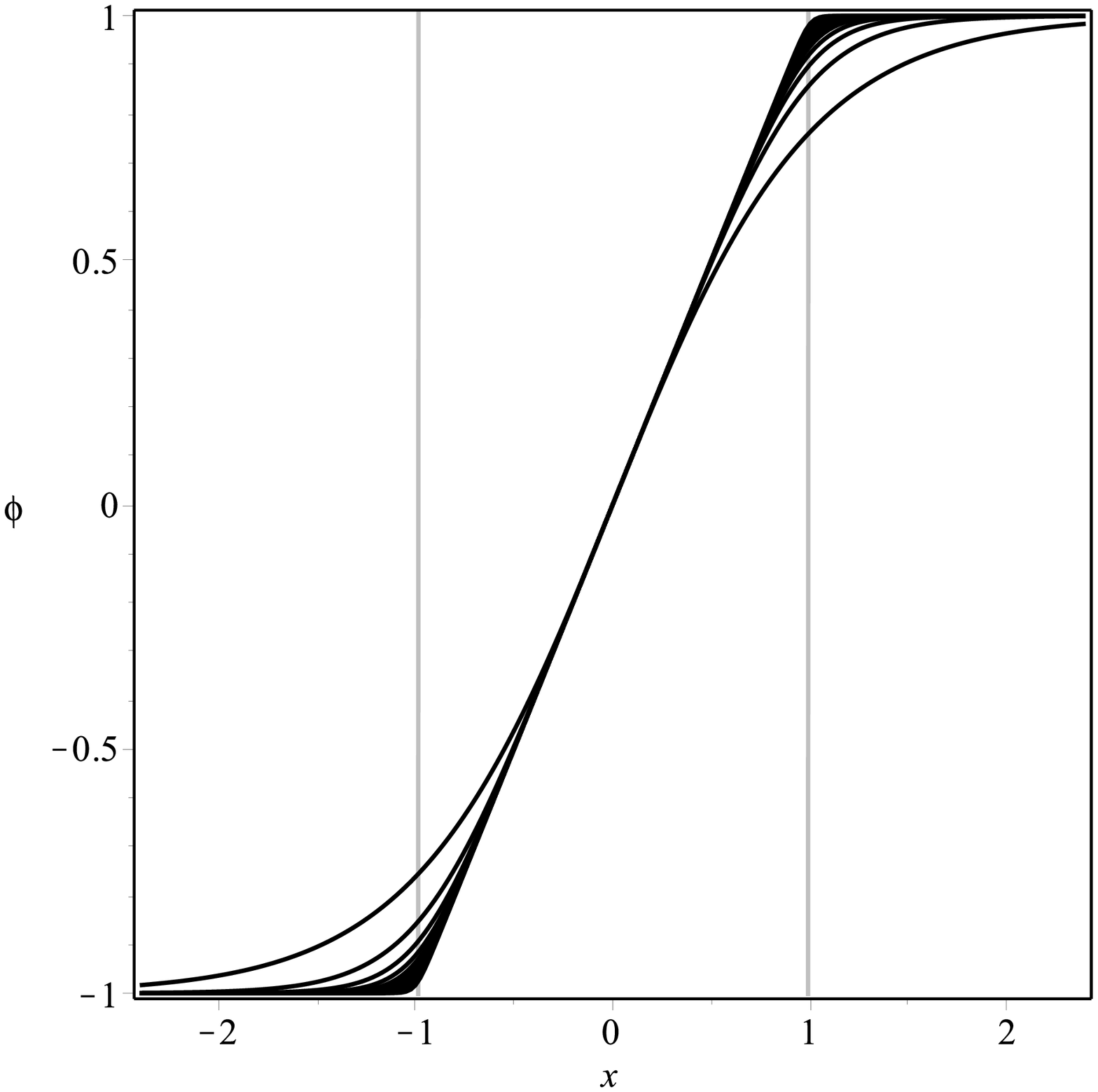}
\includegraphics[width=4.2cm]{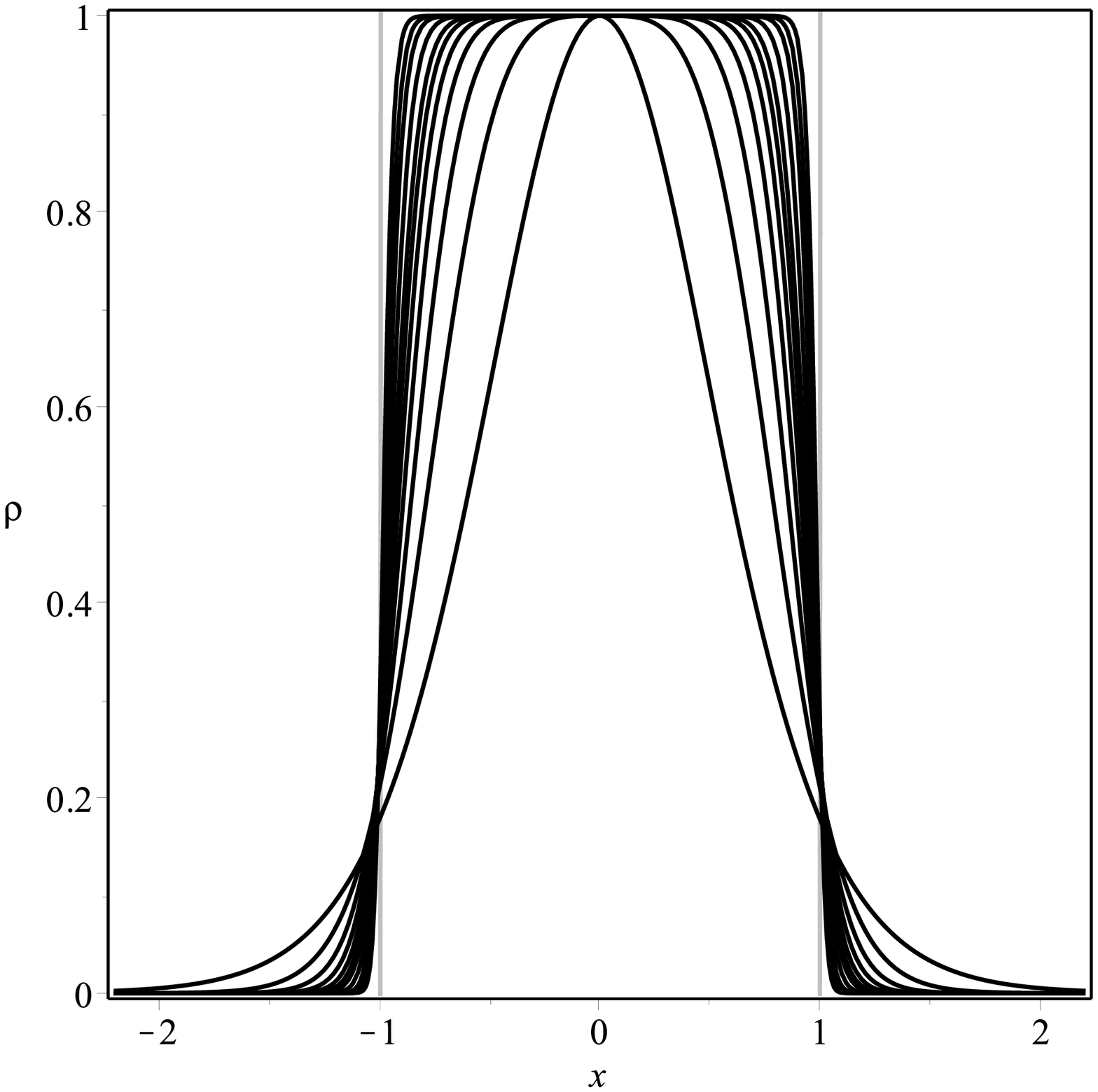}
\includegraphics[width=4.2cm]{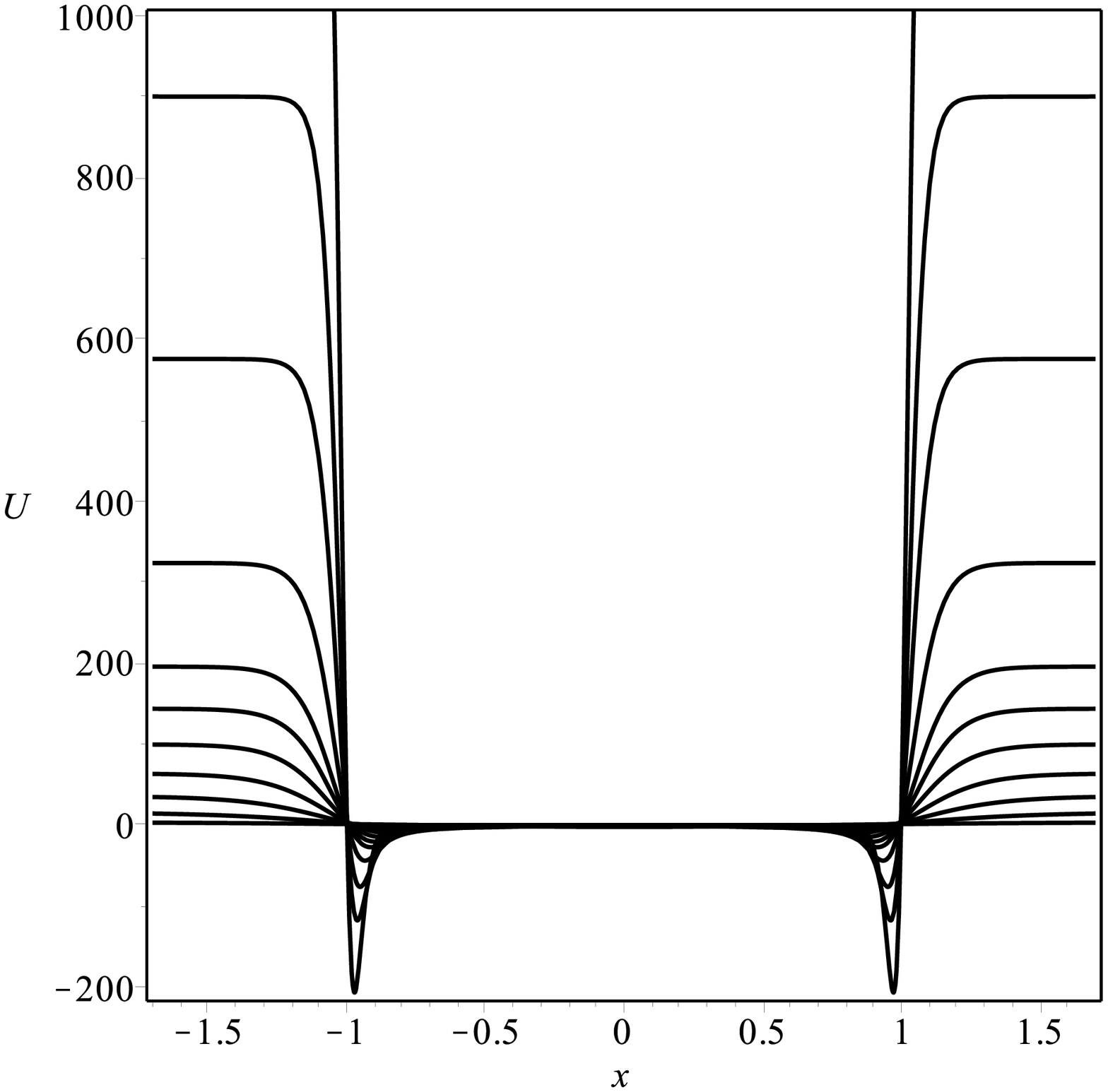}
\caption{The potential \eqref{potmodel2} (top left), its corresponding solution (top right), energy density (bottom left) and Schroedinger-like potential (bottom right), depicted for $n=1$, and increasing to larger and larger values.}
\label{fig4}
\end{figure}

Let us now consider another model, described by
\be\label{ln}
{\cal L}_n=\frac12 \partial_\mu\phi\partial^\mu\phi-V_n(\phi),
\ee
where the potential is given by
\be\label{potmodel2}
V_n(\phi)=\frac12 \left(1-\phi^{2n} \right)^2,
\ee
with $n$ being integer, $n\geq1$.  For $n=1$, we get back to the $\phi^4$ model. This model is different from the previous one, and the parameter $n$ is now integer.

\begin{figure}[t]
\includegraphics[width=7.2cm]{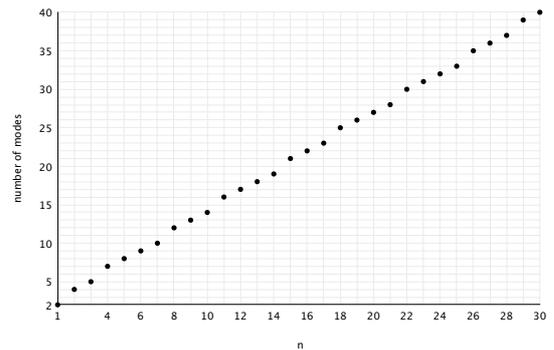}
\caption{Number of bound states of the stability potential of the second model, depicted as a function of $n$.}
\label{fig5}
\end{figure}
\begin{figure}[t]
\includegraphics[width=4.2cm]{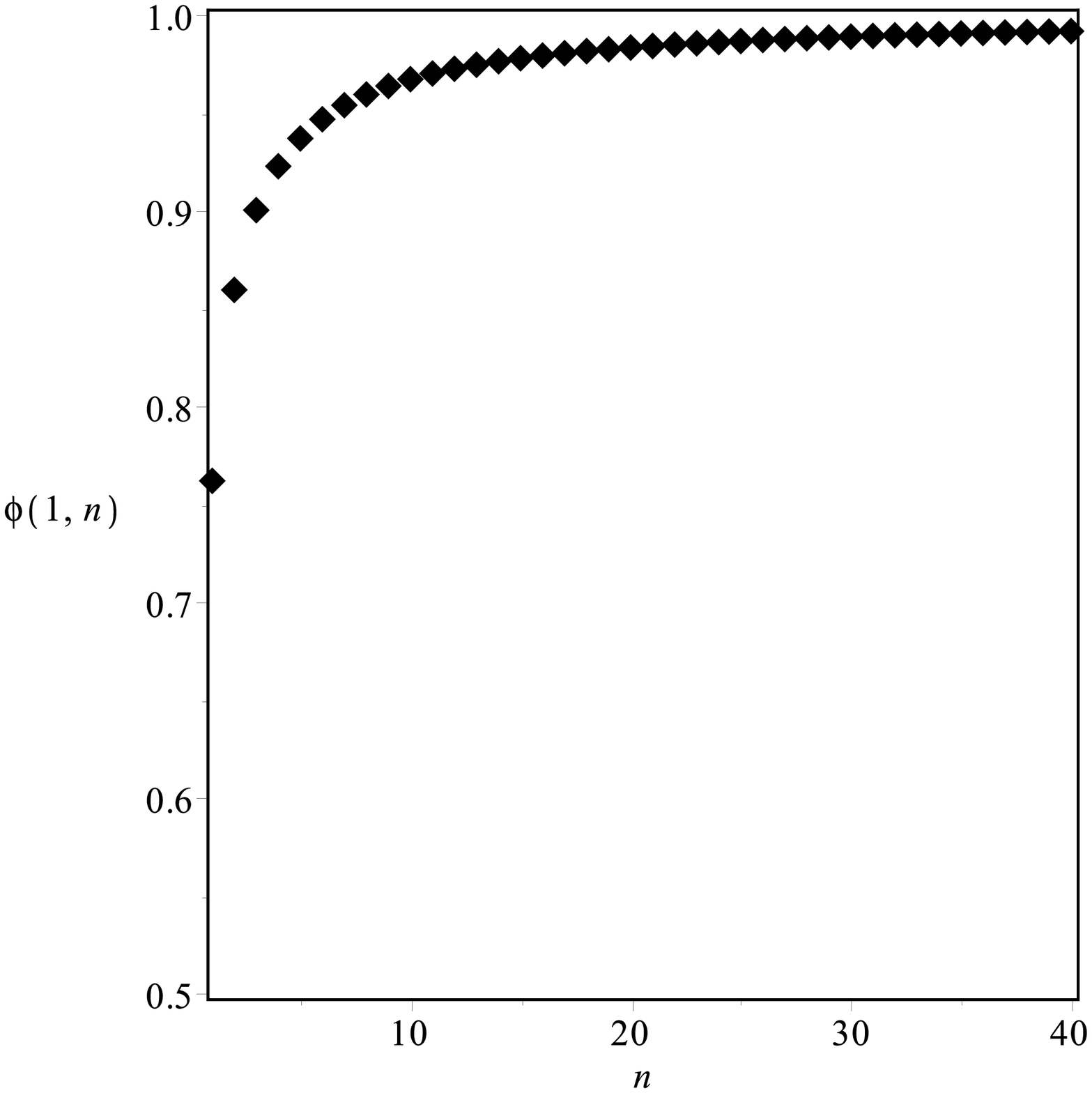}
\includegraphics[width=4.2cm]{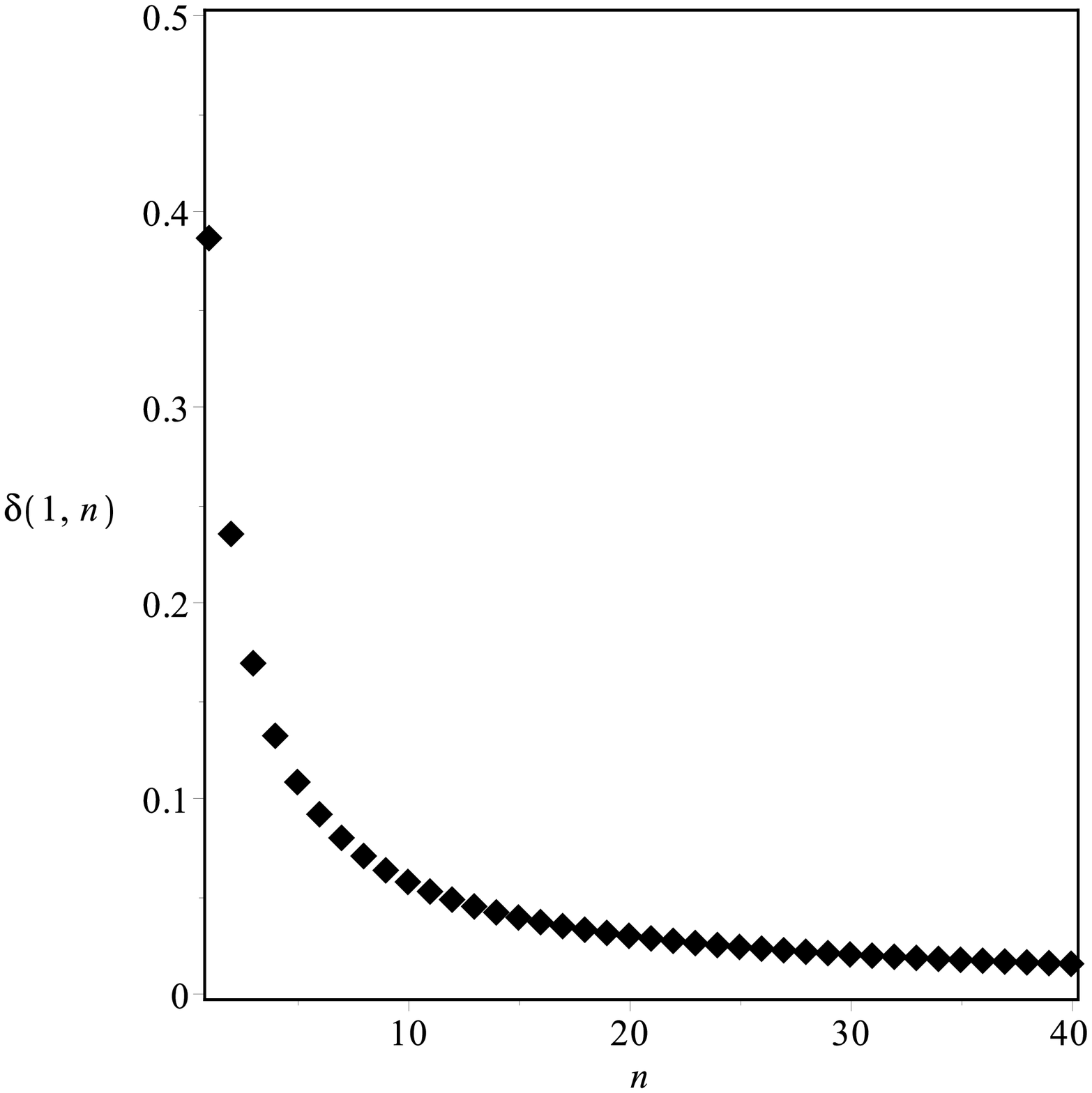}
\caption{Behavior of $\phi(1,n)$, the kink solution $\phi(x)$ calculated at $x=1$ (left) and the relative energy $\delta(1,n)$ of the kink solution (right), for several values of $n$.}
\label{fig6}
\end{figure}

The equation of motion for this model is
\be\label{eomn}
\frac{d^2\phi}{dx^2}=-2n\phi^{2n-1}(1-\phi^{2n}).
\ee
The second derivative of the potential is
\be
\frac{d^2V_n}{d\phi^2}=4n^2\phi^{4n-2} - 2n(2n-1)\phi^{2n-2}(1-\phi^{2n}).
\ee
This potential is non-negative and has the two minima $\phi_\pm=\pm1$, with $V_n(\pm1)=0$, for any $n$. Also, the mass is such that $m_n^2=4n^2$, which increases to very large values for very large values of $n$. 

We solve the equation of motion \eqref{eomn} numerically, and in Fig.~4 we depict the potential (top left), the kink solution (top right), the energy density (bottom left) and the stability potential (bottom right) for several values of $n$. 
We focus on the stability potential, to see how it changes with increasing $n$. It is depicted in Fig.~{\eqref{fig4}}, bottom right. It is a well with increasing walls that gets more and more bound states, as $n$ increases to larger and larger values. We have investigated how the number of bound states increases with increasing $n$, and we depict the results in Fig.~\eqref{fig5}.

In Fig.~6 we studied two other quantities numerically, one being $\phi(1,n)$, that is, the value of the kink solution $\phi(x)$ at the point $x=1$, as a function of $n$. The other quantity is the energy inside the interval $[-1,1]$. We studied $\delta(1,n)=1-E(1,n)/E(1,\infty)$, where  $E(1,n)$ is the quantity of energy of the kink with a given $n$, inside the interval $|x|\leq1$, and $E(1,\infty)=\lim_{n\to\infty} E(1,n)$. The results show that the kink solution tends to the unit value at $x=1$, for increasing values of $n$, and that the quantity of energy of the kink inside the compact interval $|x|\leq1$ tends to the energy $E(1,\infty)$ for larger and larger values of $n$. These results suggest that the kink tends to a compact structure for $n$ increasing to a very large value.

{\it Hybrid brane.} Let us now consider the scalar field in $AdS_5$ warped geometry with a single extra dimension of infinite extent. We follow \cite{RS,GW,MC,brane} and take the line element for flat brane in the form
\be
ds^2_5=e^{2A}\,\eta_{\mu\nu}dx^\mu dx^{\nu}-dy^2,
\ee
with $A=A(y)$ being the warp function, $\eta_{\mu\nu}$ describing the four-dimensional ($\mu,\nu=0,1,2,3$) Minkowski spacetime, and $y$ standing for the extra dimension. In this case, the Einstein-Hilbert action has the form
\be
I=\int d^4x dy\,\sqrt{|g|} \left(-\frac14 R+{\cal L(\phi,\partial_\mu\phi)}\right),
\ee
where $R$ is the scalar curvature and ${\cal L}(\phi,\partial_ \mu\phi)$ describes the scalar field, and we are using $4\pi G_{5}=1$.

\begin{figure}[t]
\includegraphics[width=4cm]{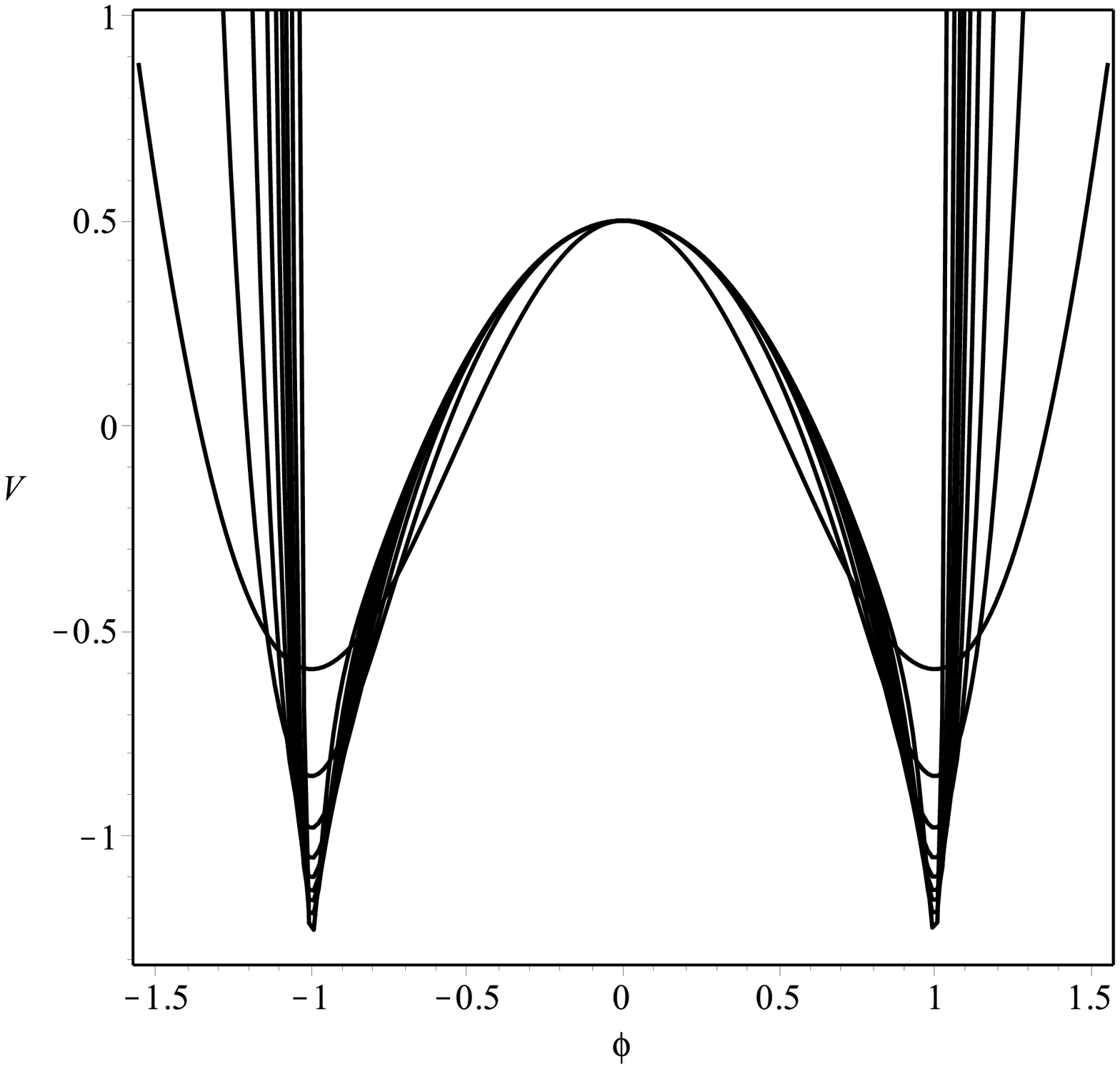}
\includegraphics[width=4cm]{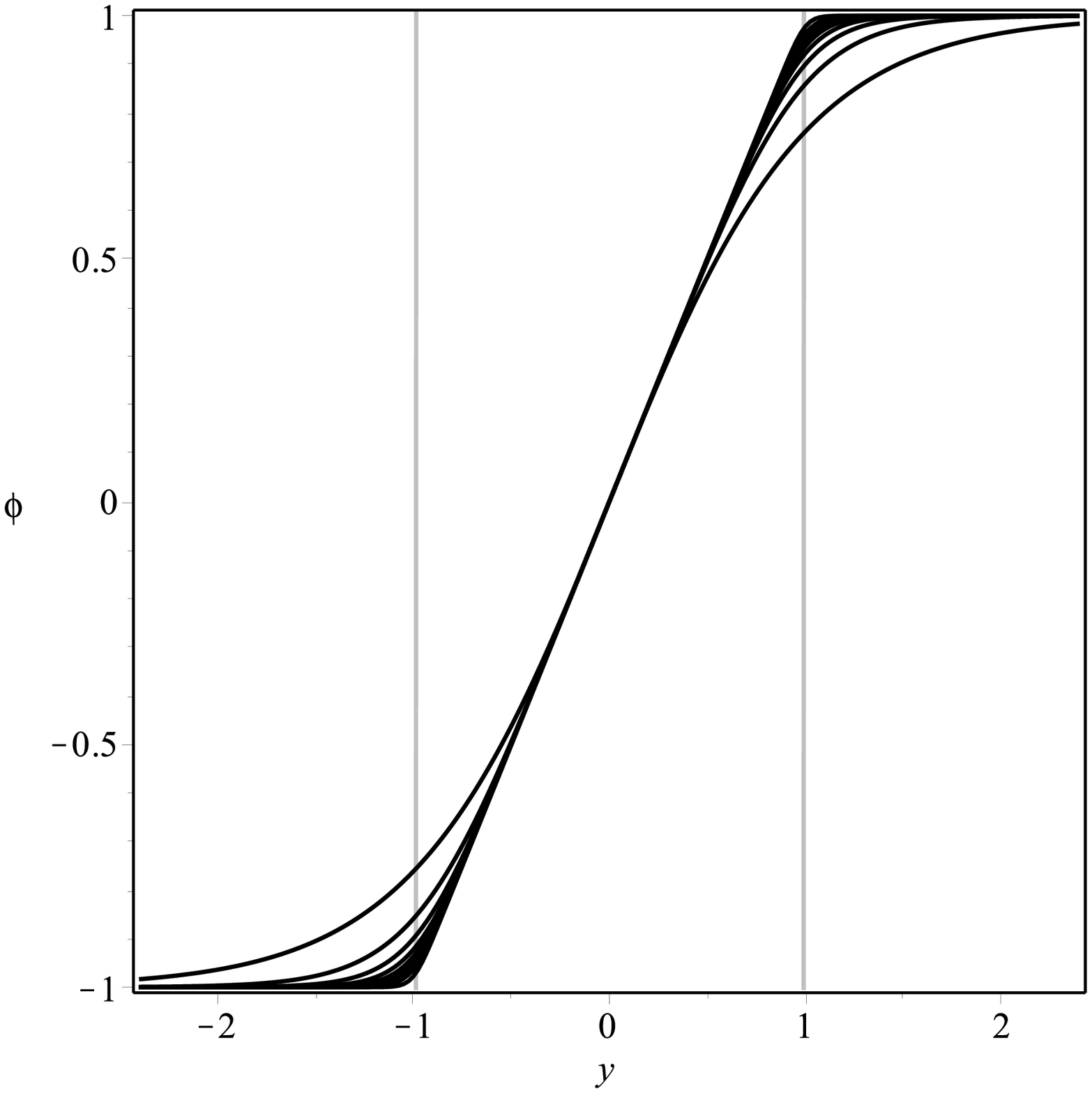}
\includegraphics[width=4cm]{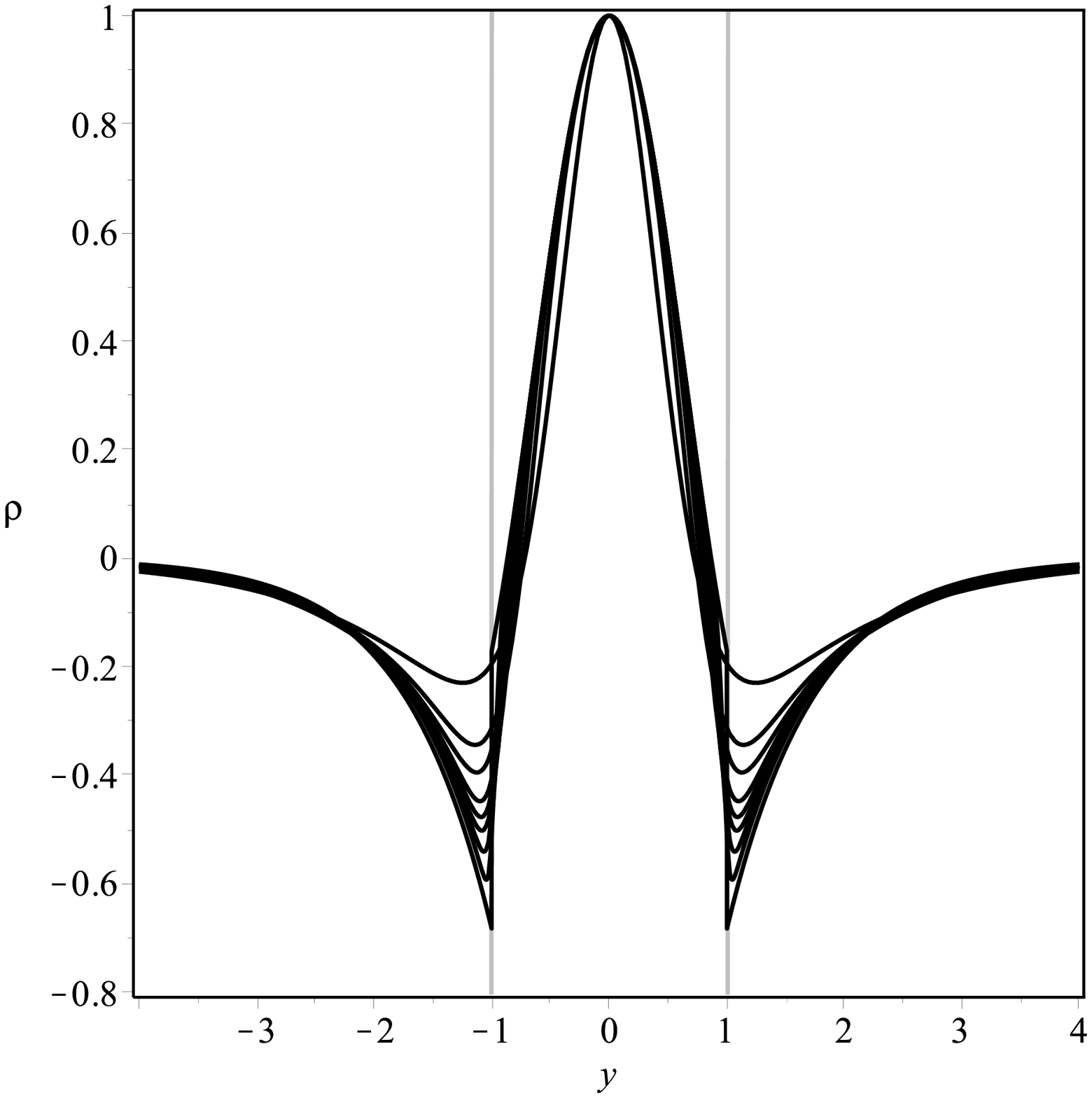}
\includegraphics[width=4cm]{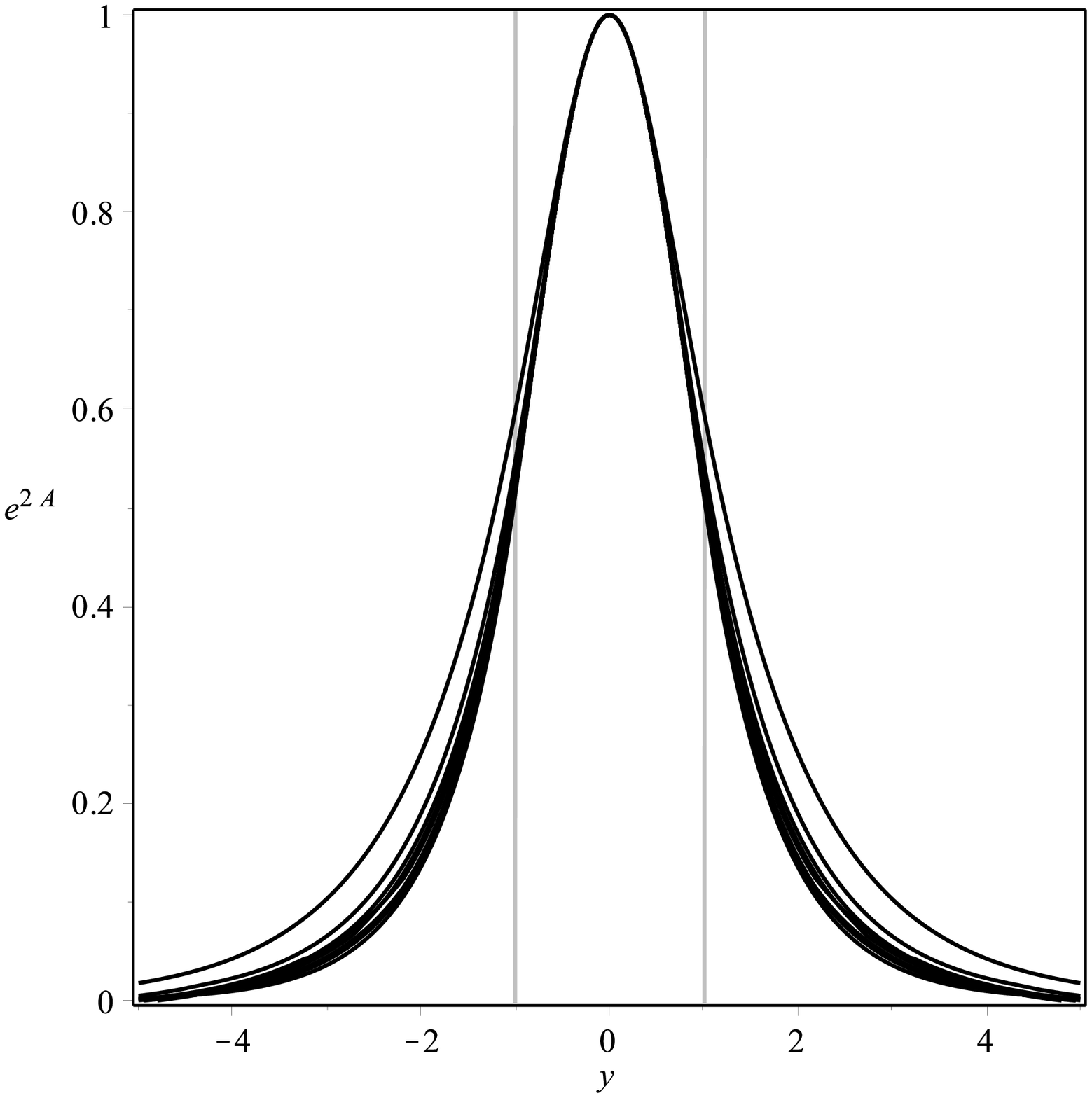}
\caption{The potential \eqref{pw} (top left), scalar field solution (top right), energy density (bottom left) and warp factor (bottom right), depicted for $n=1$, and increasing to larger and larger values.}
\label{fig7}
\end{figure}

We illustrate this case with the second model, with ${\cal L}_n$ defined by Eqs.~{\eqref{ln}} and \eqref{potmodel2}. The potential $V_n(\phi)$ can be written as
\be\label{wn}
V_n(\phi)=\frac12\left(\frac{dW_n}{d\phi}\right)^2;\;\;\;\;\;\;
W_n(\phi)=\phi-\frac{\phi^{2n+1}}{2n+1}.
\ee
We suppose that the scalar field only depends on the extra dimension, $\phi=\phi(y)$. In this case, the scalar field equation of motion has the form 
\bes\ben
\frac{d^2\phi}{dy^2}+4\frac{d\phi}{dy} \frac{dA}{dy} =\frac{dV_n}{d\phi}.
\een
Also, the several Einstein's equations reduce to 
\ben
\frac{d^2A}{dy^2}&=&-\frac23 \left(\frac{d\phi}{dy}\right)^2, \\
\left(\frac{dA}{dy}\right)^2&=& \frac{1}{6}\left(\frac{d\phi}{dy}\right)^2-\frac13 V(\phi).
\een
\ees 
We now take 
\be 
\frac{dA}{dy}=-\frac23 W(\phi);\;\;\;\;\;\; \frac{d\phi}{dy}= \frac{dW}{d\phi}.
\ee
These first-order equations solve the equations of motion if the potential is written as
\be
V(\phi)=\frac12 \left(\frac{dW}{d\phi}\right)^2 - \frac43 \,W^2(\phi).
\ee
We then use $W$ as in Eq.~\eqref{wn} to get
\be\label{pw}
V(\phi)=\frac12 (1-\phi^{2n})^2 - \frac43 \phi^2\left(1-\frac{\phi^{2n}}{2n+1} \right)^2,
\ee
which we depict in Fig.~\eqref{fig7}. This is the potential in curved spacetime; it gives $V(\pm1)=-(4/3)(2n/(2n+1))^2$, which becomes $-4/3$ for very large values of $n$.

The scalar field solutions are the same of the previous model, studied in flat spacetime; they are obtained numerically, and are also plotted in Fig.~\eqref{fig7}. Moreover, we can use the first-order equations to write the warp function in terms of the scalar field analytically, in the form 
\be
A(\phi)=-\frac13\,{\frac {{\phi}^{2}}{2\,n+1}}-\frac{2n{\phi}^{2}}3\,{\frac {\,
{\,\mbox{$_2$F$_1$}(1,\frac{1}{n};\,1+\frac{1}{n};\,{\phi^{2n}})}}{2\,n+1}},
\ee
where $_2F_1$ is hypergeometric function. Also, we can write the energy density in the form
\be
\rho =e^{2A}\left( \left(\frac{dW}{d\phi}\right)^2-\frac43\, W^2\right).
\ee
We can use the scalar field $\phi=\phi(y)$ to depict the warp factor and the energy density as functions of $y$, as we show in Fig.~\eqref{fig7}. Because the kink transmutes into a compact kink for very large values of $n$, we note that the warp factor decays as in the thin brane case, when the extra dimension goes outside the compact interval $[-1,1]$. In fact, for $n$ very large, in the limit where the kink tends to become compacton, the warp function has the form
\be
A(y)=
\begin{cases}
-\frac13 y^2,\,\,\,&|y|\leq 1;\\
-\frac23 |y|+\frac13, \,\,\, & |y|>1.
\end{cases}
\ee 
In the same limit, the warp function can be used to write the energy density analytically, as
\be
\rho(y)=e^{2A(y)}
\begin{cases}
-\frac43 y^2+ 1,\,\,\,&|y|\leq 1;\\
-\frac43 , \,\,\, & |y|>1.
\end{cases}
\ee
We see from the above expression that the energy density has a finite discontinuity; it is, however, integrable and gives zero total energy, as it happens with a standard thick brane. In fact, we have checked numerically that the energy is zero for any $n$, ranging from $n=1$, and increasing to larger and larger values. 

The braneworld profile discloses a different behavior, since we know that it is only asymptotically that a standard thick brane behaves as a thin brane. Here, however, the brane behaves as a thin brane when the extra dimension is outside the compact space $[-1,1]$. We call this hybrid configuration a hybrid brane. We have investigated linear stability of the gravity sector in this hybrid scenario, and shown that it is stable. The investigation is similar to the one done in Ref.~\cite{b}, so we omit it here.

{\it Summary.} In this work we studied kinks and compactons in relativistic models described by a single real scalar field in two-dimensional spacetime. We considered the case where the scalar field presents standard kinematics, and we could unveil a nice way to smoothly go from kinks to compactons. We considered two distinct possibilities, one controlled by a real parameter, and the other by another parameter, integer. The same methodology may be used to investigate the transition from vortices to compact vortices, and from monopoles to compact monopoles, in $(2,1)$ and in $(3,1)$ spacetime dimensions, respectively. These issues are presently under consideration.

We also investigated how the procedure works in the five-dimensional braneworld scenario, for a flat brane with a single extra dimension of infinite extent. We considered the second model ${\cal L}_n$, defined by Eqs.~\eqref{ln} and \eqref{potmodel2}, controlled by the integer $n$, and we introduced first-order differential equations that solve the equations of motion. The braneworld scenario is robust, and the limit of very large values of $n$ leads to a hybrid brane, for which the warp factor behaves as in the thick brane case, when the extra dimension varies inside the compact space $[-1,1]$, and as in the thin brane case, when the extra dimension varies outside the compact space $[-1,1]$.

We would like to thank the Brazilian agencies CAPES and CNPq for partial financial support.


\end{document}